# Design and Implementation of an Efficient Onboard Computer System for CanSat Atmosphere Monitoring

Abhijit Gadekar*

*School of Electronics and Communication Engineering, Dr. Vishwanath Karad MIT World Peace University, Pune, 411038, Maharashtra, India*

**Abstract**

Withadvancements intechnology,thesmallerversions of satelliteshavegained momentum inthespace industryforearthmonitoringand communication-based applications. The rise of CanSat technology has significantly impacted the space industry by providing a cost-effective solution for space exploration. CanSat is a simulation model of a real satellite and plays a crucial role in collecting and transmitting atmospheric data. This paper discusses the design of an Onboard Computer System for CanSat, used to study various environmental parameters by monitoring the concentrations of gases in the atmosphere. The Onboard Computer System uses GPS, accelerometer, altitude, temperature, pressure, gyroscope, magnetometer, UV radiation, and air quality sensors for atmospheric sensing. A highly efficient and low-power ESP32 microcontroller and a transceiver module are used to acquire data, facilitate seamless communication and transmit the collected data to the ground station.

## 1. Introduction

Satellites are man-made objects that orbit around the Earth and are used for various purposes. There are different types of satellites which include communication satellites, navigation satellites, weather satellites, remote sensing satellites, and scientific research satellites used for various applications. CanSat is one such type of small satellite that is designed to fit inside a can, often used for educational purposes. While CanSats share a similar design to CubeSats, they are significantly smaller and less costly to develop [1]. They are typically launched from rockets and reach altitudes of a few hundred meters before descending back to Earth using parachutes. During their flight, CanSats can collect various types of atmospheric data.

The Onboard Computer (OBC) of a CanSat is a small, autonomous computer system that plays a critical role in the success of the CanSat mission [2]. The OBC is responsible for

* Corresponding author.

*Email address:* 1032180811@mitwpu.edu.in (Abhijit Gadekar)

controlling and monitoring CanSat's various subsystems, including data acquisition, power management, communication, and sensor calibration. The OBC is designed to operate in space, where it must function reliably in the presence of radiation, extreme temperatures, and vibration. It is built using off-the-shelf components such as microcontrollers, sensors, communication devices, and custom-designed software algorithms to the specific mission requirements.

This paper presents an in-depth investigation into the design and performance of the low-cost OBC of a CanSat suitable for STEM education. The study includes a detailed analysis of the hardware components, power management system, software architecture, and cost-effectiveness of the OBC. It investigates the objectives of the OBC designed for a CanSat, which include recording the air qualityand UVradiation usingatmospheric remotesensing sensors, tracking CanSat's health parameters such as its acceleration, calibration, gyroscope, and power consumed, recording atmospheric parameters such as temperature, pressure, humidity, and altitude, obtaining exact location coordinates using GPS, and transmitting the acquired payload data through the uplink to the ground station in real-time. The low-cost OBCalsohasamodulardesignarchitecture, allowingcomponentssuchassensorsandcommunication devices to be easily replaced or upgraded, providing flexibility in educational applications.

Theremainingsectionsofthispaperare organizedasfollows. Section 2 providesaliterature review of the CanSats forvarious applications. Section 3 offers a detailed description of the design of an Onboard Computer System (OBC), highlighting the key components such asthehardwaresubsystem, groundstation, andsoftwaresubsystem. It providesathorough overview of the design process, including the development of each subsystem, which is integrated to achieve optimal performance for CanSat. The operational functioning of OBC is described in detail in Section 4. Section 5 presents the results of the OBC performance and discussions of the findings. Section 6 summarizes the key findings from the study and a conclusion on the design of an OBC for a CanSat, as well as its potential applications. Section 7 offers recommendations that include suggestions for further improvements.

## 2. Materials and Methods

In recent years, CanSats have gained popularity as an educational tool for teaching students about aerospace engineering, electronics, and other related subjects. They are relatively low-cost, which makes them accessible to a wide range of students and educators. A CanSat OBCwithautonomouscontrolandwirelesscommunicationcapabilitiesweredeveloped for an educational satellite system [3]. The OBC performs functions such as attitude control, wireless data transmission, temperature and humidity sensing. It uses an Arduino microcontroller and a Zigbee wireless communication module to establish communication between the OBC and the ground station. An autonomous control algorithm was deployed, which enables the CanSat to perform tasks such as detecting and correcting the orientation during flight.

Thesoftwareframeworkof CanSatcanbedevelopedwiththereal-timeoperatingsystem (RTOS) using the FreeRTOS kernel [4]. The development of RTOS allowed for reliable and efficient execution of the software on the OBC. An Atmel AVR microcontroller, specifically the ATmega328P, was used as the brain for the OBC. A range of sensors was used for atmospheric monitoring, such as a GPS receiver, a temperature sensor, an accelerometer, and a radio transceiver for wireless communication. The OBC was made to be compact and lightweight, weighing only 60 grams and consuming minimal power. The performance of an OBC was evaluated in terms of its power consumption, data acquisition and processing capabilities, and wireless communication range.

A miniature OBC for CanSat was implemented using commercial-off-the-shelf (COTS) components. The use of the Raspberry Pi as the CPU, combined with other COTS components and open-source software, provided a practical and efficient solution for controlling and communicating with the CanSat [5]. Another low-cost OBC was developed for a CanSat using AVR microcontroller ATmega2560. The OBC includes a real-time clock (RTC) module for timekeeping and a micro SD card for data storage and is programmed in the C language using the Atmel Studio Integrated Development Environment (IDE) [6].

A modular OBC based on an ARM Cortex-M3 microcontroller was developed, capable of handling various tasks such as data acquisition, telemetry, and control of the CanSat. It allows for easy customization and expansion [7]. The development of a Raspberry Pi OBC equipped with sensors for measuring temperature, humidity, pressure, and acceleration was accomplished for monitoring the atmosphere. It was capable of transmitting data to a ground station using a radio module [8]. A low-cost OBC based on an Arduino microcontroller was designedforhandlingtaskssuchasdataacquisition,telemetry,andcontroloftheCanSat[9].

## 3. Theory and Calculations

Atmospheric parameters are variables that describe the state of the Earth's atmosphere. They include physical properties such as temperature, pressure, humidity, and wind speed, as well as chemical properties such as greenhouse gas concentrations and air quality. The rise in air pollution levels can be attributed to multiple factors including increased road transportation vehicles, industries, and other sources. However, accurate measurement of air pollutants is hindered by the high cost of environmental monitoring processes, leading to a lack of reliable data on the extent of the problem. Measuring atmospheric parameters is essential forunderstanding weather patterns, climate change, and air qualityand making predictions.

The proposed Onboard Computer System for a CanSat focuses on achieving the objectives of recording air quality and UV radiation, tracking CanSat's health parameters, recording atmospheric parameters, obtaining the exact location coordinates of the CanSat, and transmitting the acquired payload data in real-time to the Ground Station. The Onboard Computer System is divided into three systems: The hardware system of CanSat, the Ground Station System, and the Software system of CanSat. Figure 1 below shows system block diagram of OBC implemented for a CanSat.

### 3.1. Hardware System of CanSat

Thehardwaredesignoftheonboardcomputersystemisacriticalcomponentofa CanSat. It is used for processing data from the CanSat's sensors, communicating with the ground station, and executing mission-specific commands and must adhere to specific requirements and limitations to ensure its successful operation. First, all components of the CanSat must be able to fit within a standard-sized can, except for the parachute. Additionally, the CanSat must be powered by a battery, providing the necessary energy for its functions. Moreover, the CanSat's weight should be kept to a minimum to allow for greater altitude and to ensure accurate measurement of various parameters at different altitude levels. Forprecise results in CanSat operations, it is crucial to consider the design constraints and requirements. The OBC's hardware system is categorized into Sensor, Communication, and Electric power subsystems which are explained in further sections.

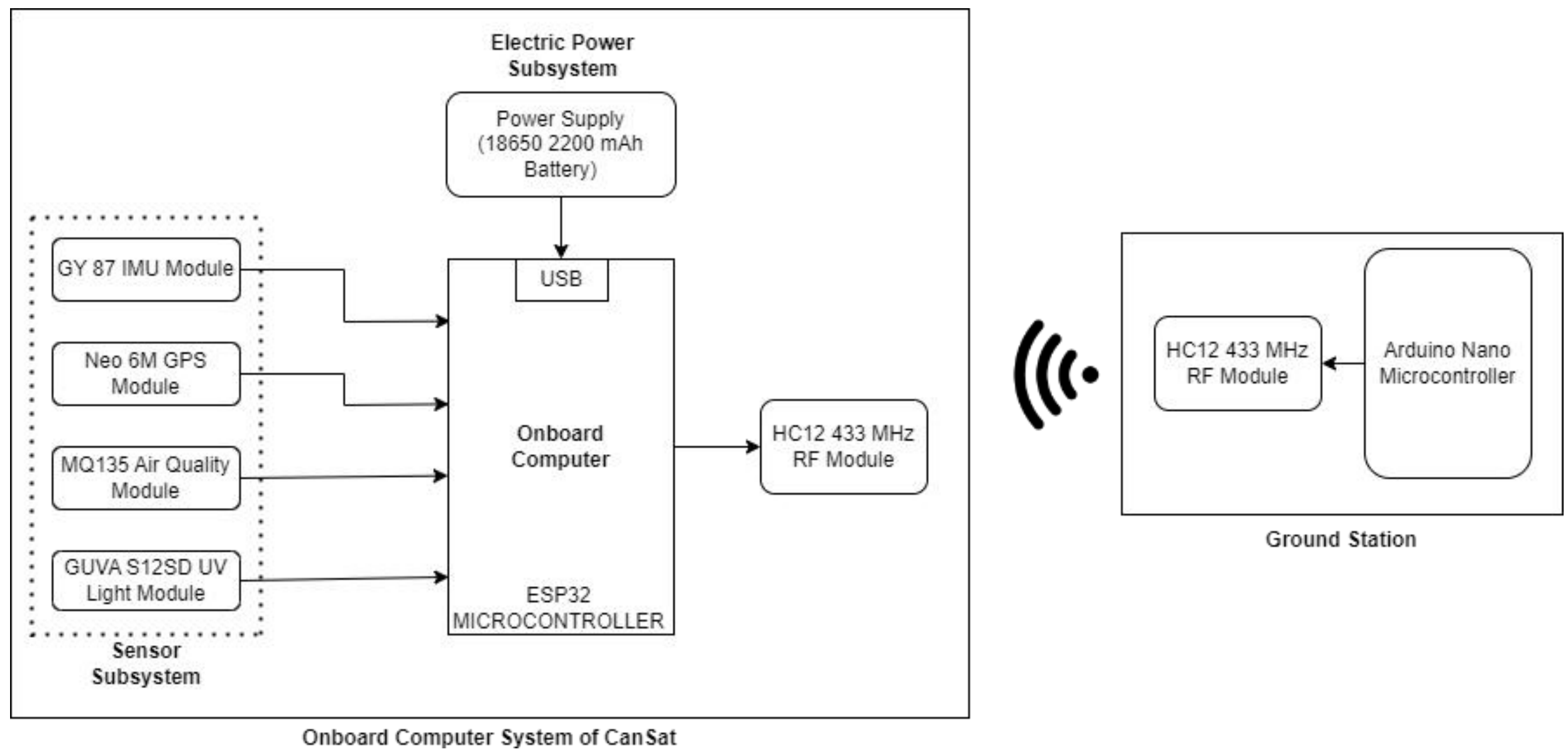

Figure 1: System Block Diagram of an OBC for a CanSat

Since OBC serves as the brain of CanSat, it is essential to consider factors such as performance, power consumption, communication capabilities, weight, and ease of use when selecting OBC hardware. Microcontrollers such as Arduino, ESP, and Raspberry Pi are popular choices for their affordability and ease of use. They are also highly customizable, allowing users to add various sensors and peripherals to their CanSat system. Table 1 presentsacomparativeanalysisconducted toselect thebest onboardcomputerfora CanSat project. ESP32 was selected as the OBC due to its low-power applications, built-in Wi-Fi and Bluetooth connectivity, and ease of programming, thus providing a cost-effective solution.

### *3.1.1. Sensor Subsystem Design*

A sensor subsystem is a set of sensors used to collect data about a specific environment orsystem. In this case, the subsystemconsists of sensors that measure different parameters such as UV radiation, altitude, temperature, humidity, pressure, GPS, acceleration, gyroscope, and air quality. In a comparison of various sensors, factors such as accuracy, range,

Table 1: Comparative Analysis of OBCs for CanSat

| Parameter | ESP32 | STM32 | Arduino Nano | Raspberry Pi Pico |
|---|---|---|---|---|
| Processor | Dual-core Tensilica LX6 32-bit | ARM Cortex-M4 32-bit | AVR ATmega328P 8-bit | ARM Cortex-M0+ 32-bit |
| Clock speed | Up to 240 MHz | Up to 180 MHz | 16 MHz | Up to 133 MHz |
| Memory | 520 KB SRAM, 4 MB flash | 128 KB SRAM, 512 KB flash | 2 KB SRAM, 32 KB flash | 264 KB SRAM, 2 MB flash |
| I/O | 34 programmable GPIO pins | 70-114 GPIO pins | 14 digital I/O pins, 6 analog pins | 26 multi-function GPIO pins |
| Connectivity | Wi-Fi, Bluetooth, I2C, SPI, UART | USB, CAN, Ethernet, I2C, SPI, UART | USB, I2C, SPI | USB, I2C, SPI |
| Operating Voltage | 2.2 V to 3.6 V, 80mA | 1.7 V to 3.6 V | 5V | 1.8 V to 5.5 V |
| Price | $4 to $8 | $5 to $20 | $5 to $15 | $4 |

sensitivity, response time, and cost were evaluated.

Instead of using individual sensors for acceleration, gyroscope, temperature, humidity, pressure, and altitude measurements, an all-in-one sensor module was chosen for the OBC. The GY-87 module includes a barometer, magnetometer, gyroscope, accelerometer, and thermometer, all on one board. This not only reduces the overhead of receiving valuesfrom different sensors, but it is also lightweight and takes up less space, making it an ideal choice for the CanSat's payload.

The GY-87 module is a single-board solution that includes multiple sensors such as the MPU6050 3-axis accelerometer and 3-axis gyroscope, HMC5883L triple-axis magnetometer, and BMP180 barometric pressure sensor. Toconnect the GY-87 moduletothe ESP32 MCU, the SDA (Serial Data Pin) and SCL (Serial Clock Pin) of the module need to be connected to the corresponding pins on the MCU. This is done via I2C communication and allows the MCU to receive sensor data from the GY-87 module.

The Neo 6M GPS module was chosen as the GPS sensor for the CanSat project due to its ability to provide accurate location information in terms of latitude and longitude. This module utilizes a combination of GPS and GLONASS satellites to determine location and is capable of providing a 10-meter accuracy. The module is also relatively small and lightweight, making it an ideal choice. The Neo 6M GPS module communicates with the MCU via UART communication.

Air quality monitoring is an essential aspect of the CanSat project, and the MQ-135 gas sensor was selected as the primary sensor for this purpose. This sensor is capable of detecting a range of gases, including carbon dioxide, ammonia, and benzene, and provides an analog output that can be read by the MCU. The MQ-135 gas sensor is designed to detect a range of polluting gases in the air. It utilizes SnO2, which has a higher resistance in clear air, as a gas sensing material. When polluting gases are present, the resistance of the sensor decreases in proportion to the concentration of the gas. The MQ-135 gas sensor is capable of detecting concentrations of up to 2000 ppm.

The GUVA-S12SD UV Light Sensor Module was chosen as the primary UV light sensor. This analog sensor is capable of converting UV light intensity into a proportional voltage output that can be measured by the MCU. The sensor communicates with the MCU via an analog input pin. It can detect ultraviolet radiation in the range of 200-400 nm. It is commonly used for detecting sunlight and measuring UV exposure. Figure 2 shows the sensor's subsystem data to the ESP32 MCU.

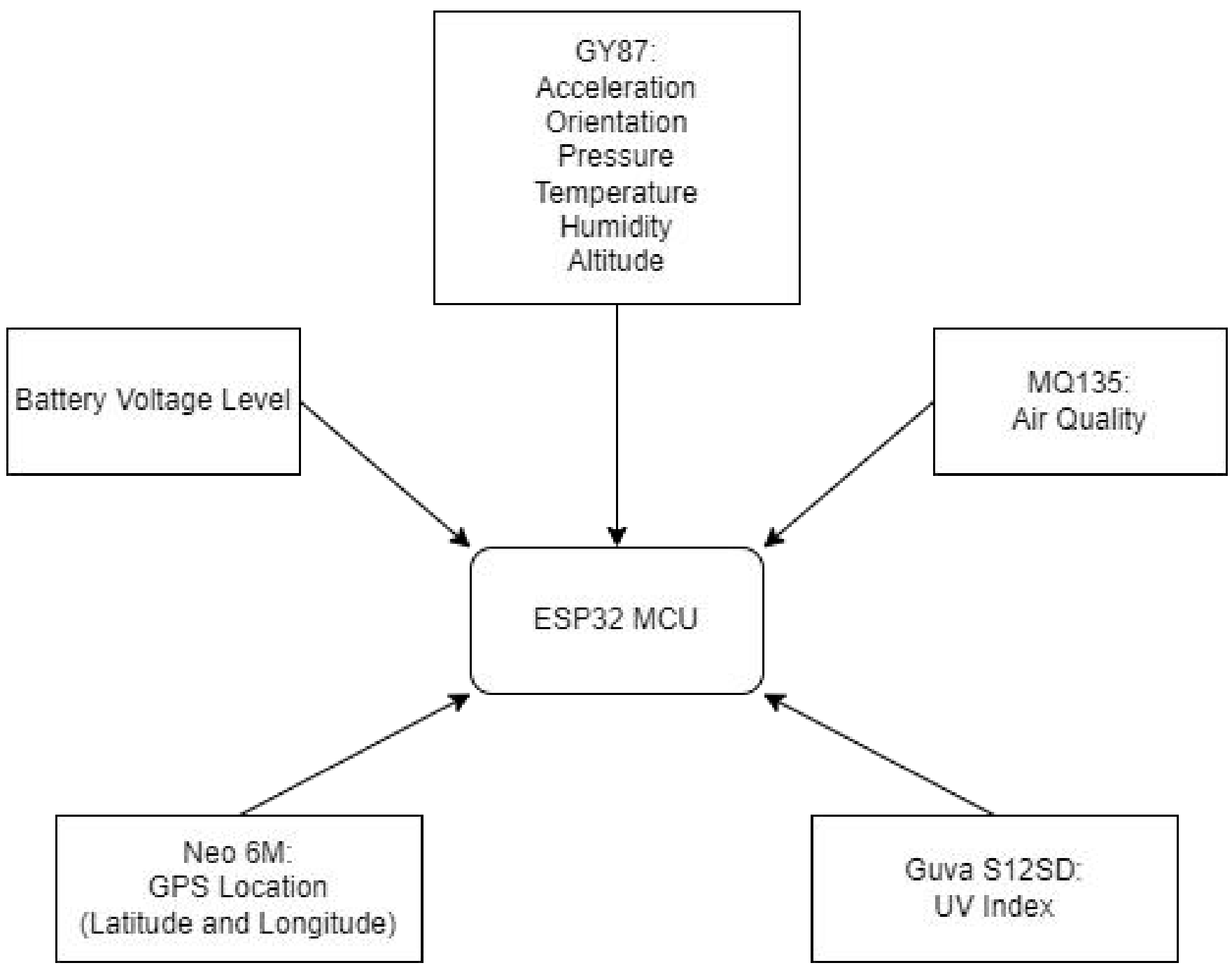

Figure 2: Sensor Subsystem Data

*3.1.2. Communication Subsystem Design*

The Communication Subsystem Design for a CanSat mission is an essential component that enables the transmission of data from the CanSat to the Ground Station. In this design, the HC12 433 MHz Transceiver Module is used as a communication device between the CanSat and the Ground Station.

The HC12 433 MHz Transceiver Module is a radio frequency (RF) module that operates in the 433 MHz frequency band. This module is chosen for its reliable performance, low powerconsumption, andabilitytooperateoveralongrange. Themoduleisequippedwithan onboardantenna, whichenablesthetransmissionandreceptionofdataoveradistanceofup to 1 km in open space. The module operates through UART serial communication protocol interfaced with the microcontroller. The HC12 433 MHz Transceiver is also equipped with several features that make it easy to use and integrate into the CanSat's Communication Subsystem. These featuresinclude a configurabledata rate, adjustabletransmit power, and an onboard status LED that indicates the module's operational status. The key parameters of HC12 module is described in Table 2 .

Table 2: Key Parameters of HC12 Communication Module

| Parameter | Value |
|---|---|
| Frequency | 433.4 - 473.0 MHz |
| Baud Rate | 1200, 2400, 4800, 9600, 19200, 38400, 57600, 115200 bps |
| Transmit Power | 0.8 - 100 mW (adjustable) |
| Sensitivity | -117 dBm |
| Operating Voltage and Operating Current | 3.2 - 5.5 VDC; 30 mA (transmitting), 19 mA (receiving), 2.5 uA (sleep mode) |

*3.1.3. Electric Power Subsystem Design*

The OBC is included with several sensors, each with its power requirements. To calculate the power needed, the operating voltage and current of each sensor were taken into consideration. This information was used to design and implement the electric power subsystem for the CanSat, ensuring that it had sufficient power to operate all of its sensors and components. The power *(P)* is product of voltage *(V)* and current *(I)* as shown in Equation 1. Table 3 shows the power budget analysis for OBC.

$$P = V \times I \qquad (1)$$

Table 3: Power Budget Analysis

| Component | Function | Voltage (Max) | Current | Operational Power |
|---|---|---|---|---|
| GY 87 | Sensor | 3.3 V | 4.034 mA | 13.315 mW |
| Neo 6M | Sensor | 3.3 V | 67 mA | 221.1 mW |
| MQ135 | Sensor | 5 V | 20 mA | 100 mW |
| Govu S12SD | Sensor | 5 V | 20 mA | 100 mW |
| HC-12 | Communication | 5 V | 150 mA | 750 mW |
| Microcontroller | ESP32 | 5 V | 250 mA | 1250 mW |
| **Total** | | | **511 mA** | |

An Orange ISR 18650 battery with an operating voltage of 3.7 V and a capacity of 2200 mAh was chosen for the requirements. Lithium Ion was chosen over Lithium Polymer due to its high leakage protection and reliability. The chosen battery was integrated into the CanSat's electric power subsystem to ensure that it had sufficient power to operate throughout the duration of the mission.

### *3.2. Ground Station Design*

Ground station design plays a significant role in receiving and processing data transmitted by the CanSat. Arduino Nano, a popular microcontroller with a wide range of I/O options and programming capabilities, was selected as the Ground Station computer. It is compact and affordable, thus making it a suitable choice for building a ground station. HC-12 module, on the other hand, is a wireless communication module that offers long-range communication capabilities at a low cost.

The HC-12 wireless transceiver module was interfaced with the Arduino Nano microcontroller board using a UART interface. The HC-12 module was connected to the Arduino's RX and TX pins, while the baud rate was set to 9600 bits per second. This allowed for the wireless transmission of data betweenthe CanSat and ground stationusing the HC-12 module, which had a range of up to 1 km in open air conditions. Figure 3 shows the interfacing diagram of HC-12 Module and Arduino Nano for Ground Station design.

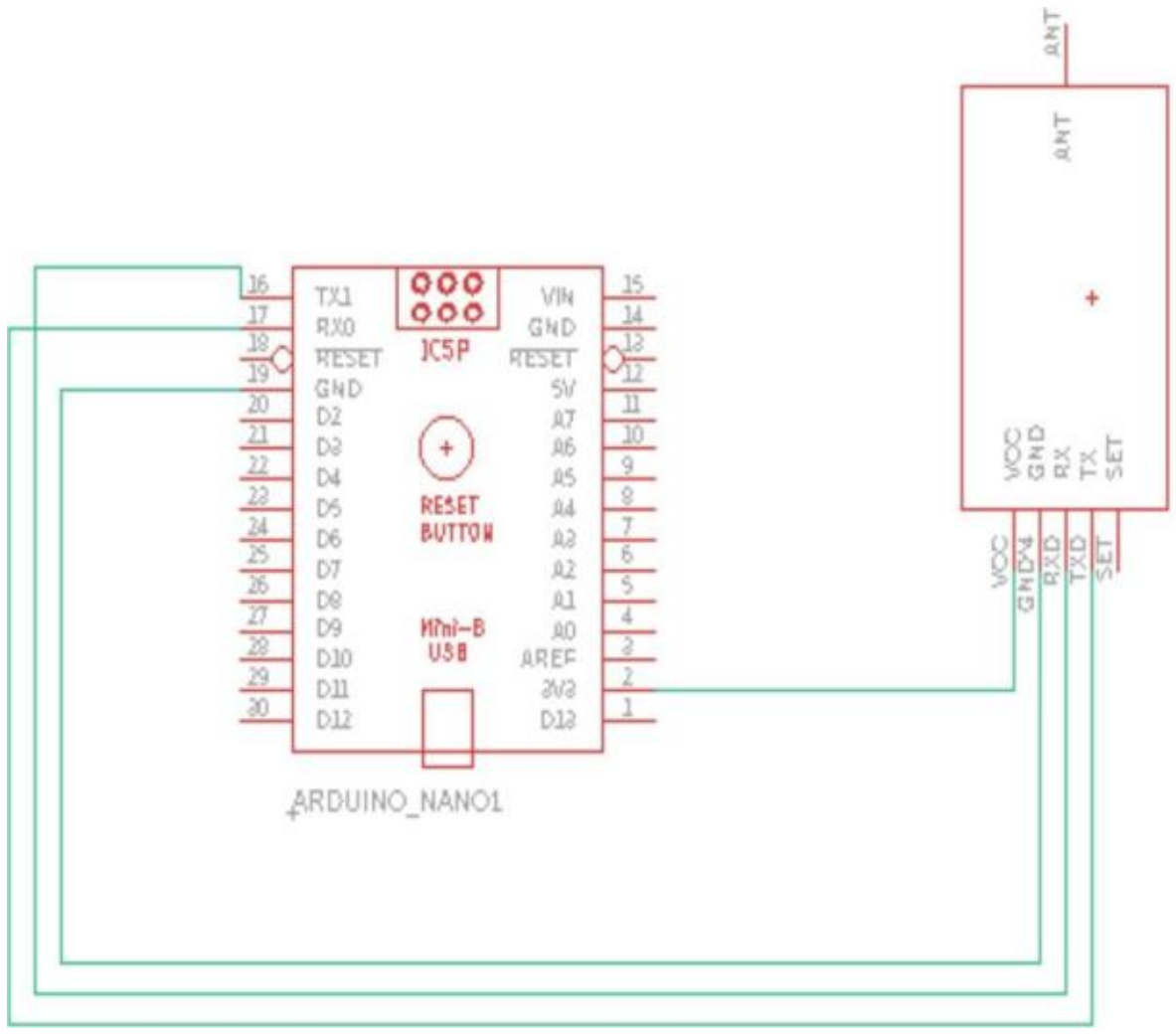

Figure 3: Interfacing Diagram of Ground Station Design

### *3.3. Software System of CanSat*

The software development plan for the CanSat OBC is categorized into three main phases. In the first phase, component-level development and testing were conducted. This phase ensured that each component of the OBC was functioning correctly before being integrated with other subsystems. The second phase, integration testing, ensured that the software was ready for final testing with integrated subsystems. The final phase, final calibration, and system testing were conducted onthe OBC and involvedtheimplementation of a web server. This phase served as the final check to ensure that the OBC was functioning correctly and met all requirements. A web server was developed to showcase sensor data retrieved from an HC12 module.

The software subsystem is developed using the Arduino Integrated Development Environment (IDE) and several libraries of MPU6050 accelerometer, BMP180 Temperature/Pressure sensor, and Adafruit NEOGPS, etc. The sensors are connected to the ESP32 microcontroller using serial communication protocols, such as I2C and UART. Figure 4 shows the flow of data collection from sensors.

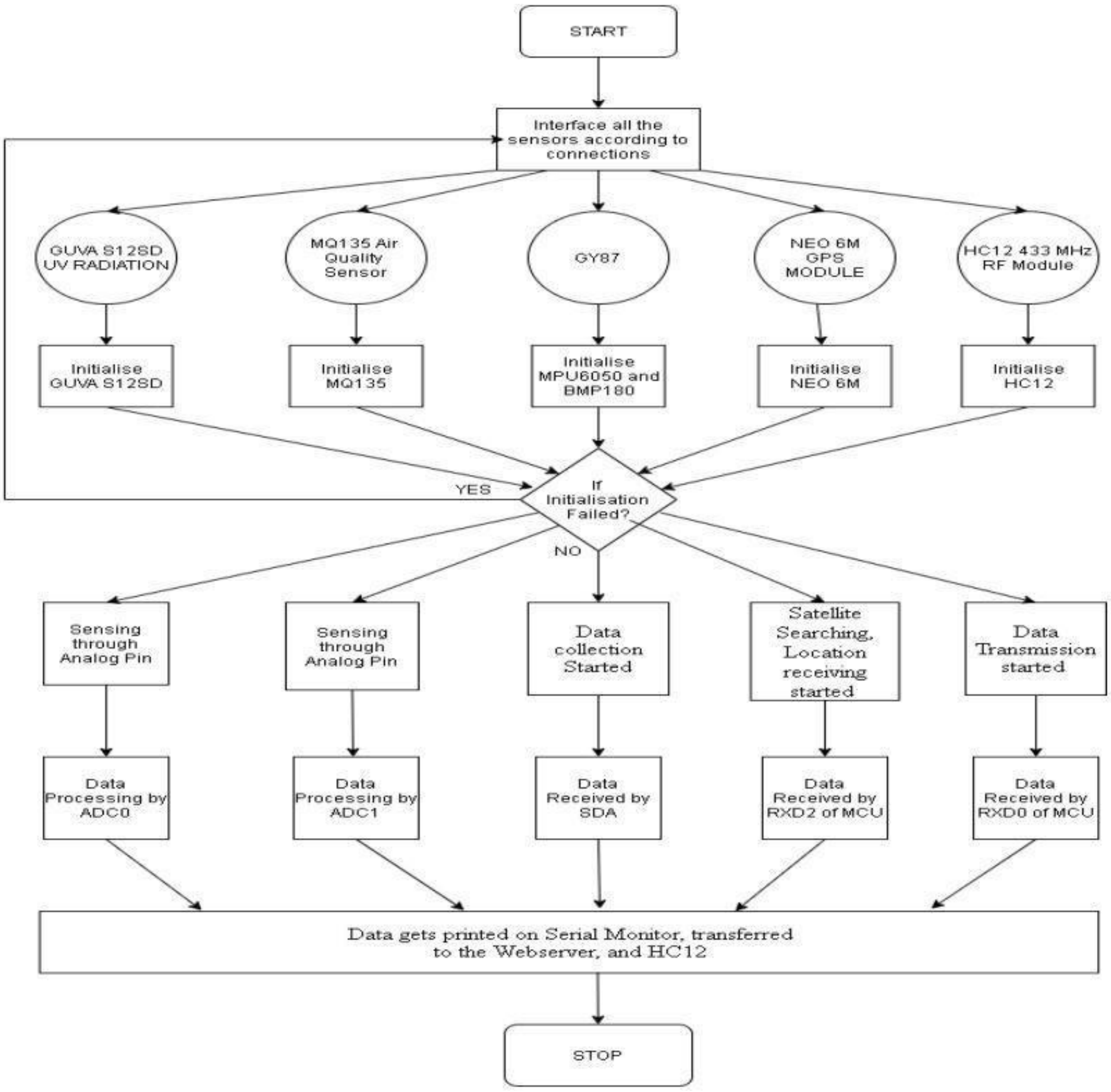

Figure 4: Flowchart of data collection

## 4. Working

The working of the designed OBC is as follows:

1) Data Acquisition: The software subsystem of the CanSat OBC acquires data from various sensors such as MQ135, NEO 6M, GUVA-S12SD, GY87, and BME280 interfaced as shown inFigure 5. The sensors measure environmental parameters such as airquality, temperature, humidity, GPS location, pressure, altitude, and UV radiation.

2) Data Processing and Transmission: The acquired data is processed by the microcontroller (ESP32) of the CanSat OBC. The microcontroller calibrates the data to ensure accuracy. The processed data is then transmitted to the ground station using wireless communication through the HC12 module. The transmission protocol includes data packaging and error checking to ensure data integrity during transmission.

3) Data Reception and Analysis: The ground station receives the transmitted data and stores it in a database for further analysis. Various statistical and analytical methods are used to extract useful information from the data.

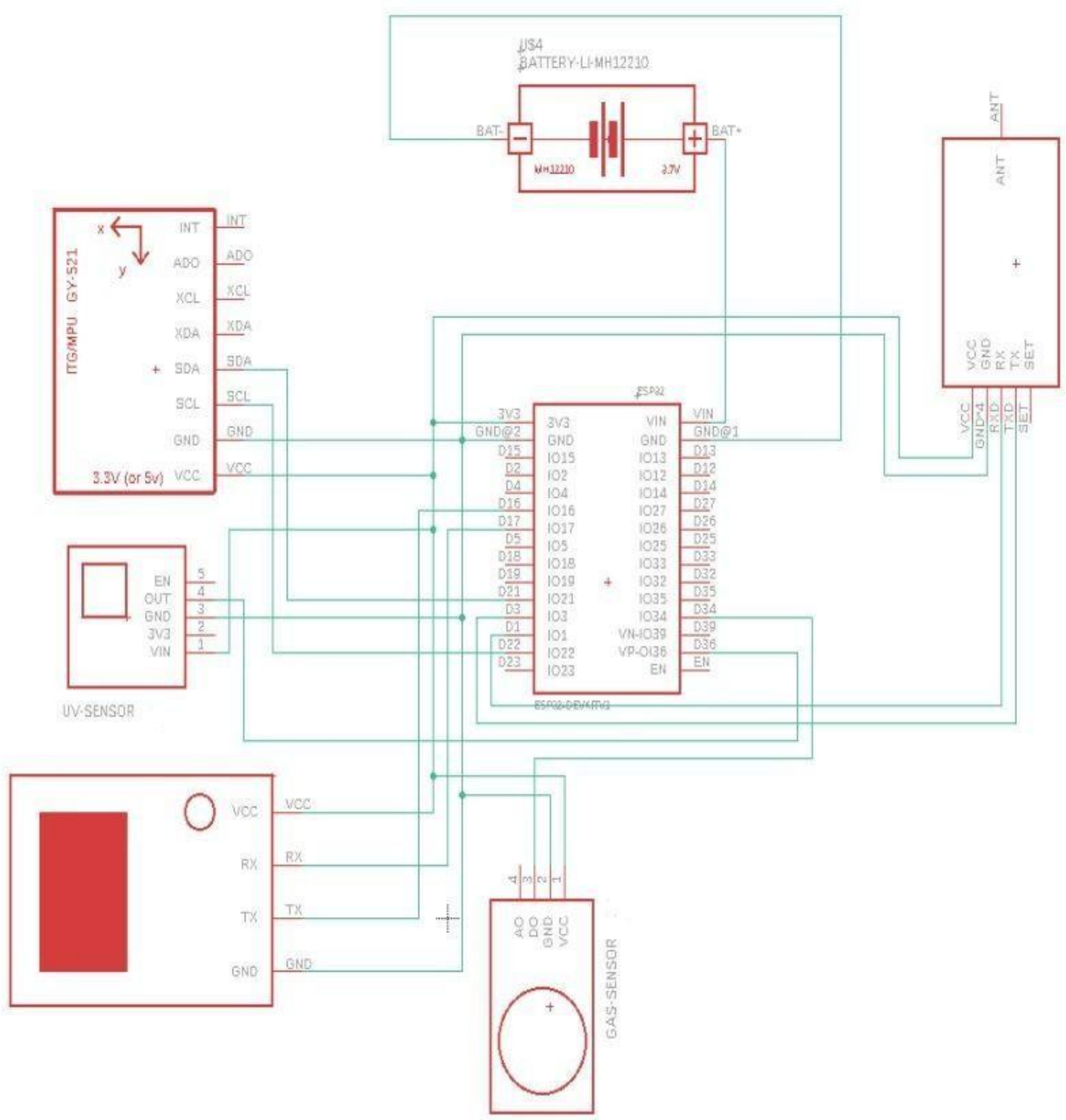

Figure 5: Interfacing Diagram of OBC

A CAD model was created to place OBC components in a CanSat, taking into account clearance, signal routing, and thermal considerations, and complying with manufacturing and performance requirements as shown in Figure 6.

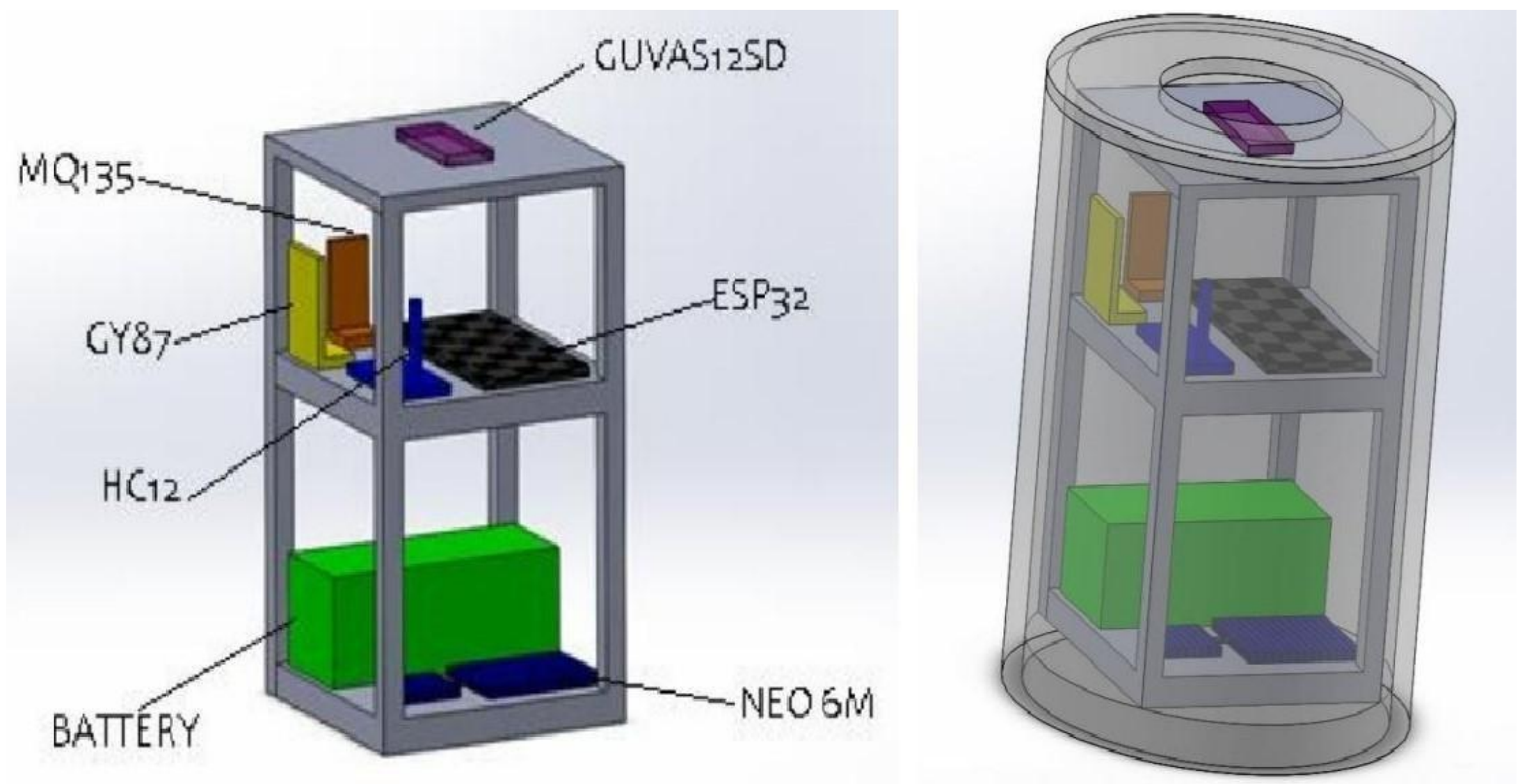

Figure 6: CAD Model of an OBC for CanSat

## 5. Results and Discussion

The CanSat OBC demonstrated effective performance duringexperimentationand testing, aligning with its theoretical calculations. Figure 7 shows the integration testing of OBC on a breadboard.

The OBC system wastestedin outdoorenvironments fromaheight rangingfromground levelupto 60 meters. The serialcommunicationbetweenthe groundstation andthe CanSat up to 800 m was accomplished by HC-12 module. The use of a web server for data from the CanSat enabled real-time monitoring of the CanSat's position and surroundings. The sensors took a calibration time of 2 seconds for accurate readings. The GUVAS12SD sensor detected UVraysinthesunlightofwavelength200-370nmandaresponsetimeoffewerthan 0.5 seconds. It was observed that the amount of UV radiation that reaches us is influenced by several factors, including the amount of cloud cover and air pollution. Thus, UVIndex is higher at high altitudes than at ground level.

The GY87 module measured acceleration in all three axes and the orientation of OBC withasensitivityof 2gand 0.1°. Thesensormeasuredtemperatureatdifferentaltitudeswith an accuracy of 1°C. It was observed that temperatures lower with altitude due to a decrease in air pressure. The BMP180 sensor on GY 87 measured atmospheric pressure in millibars which was used to compute the estimated altitude level. The GPS sensor was capable of determining the precise location of the OBC with an accuracy of up to 2 meters. The MQ135 sensor measured the Air Quality in PPM which gave vital information regarding its surroundings. The ESP32 microcontroller has an onboard analog-to-digital converter (ADC) that was used to measure its input voltage thus giving OBC's power consumption details. The battery provided a run time of 40 minutes. Figure 8 shows the data received from OBC to the ground station.

Figure 9 shows the OBC's data which is displayedon webserver after receiving toground station. The changes in atmospheric parameters with respect to different altitude levels is shown in Figure 10 .

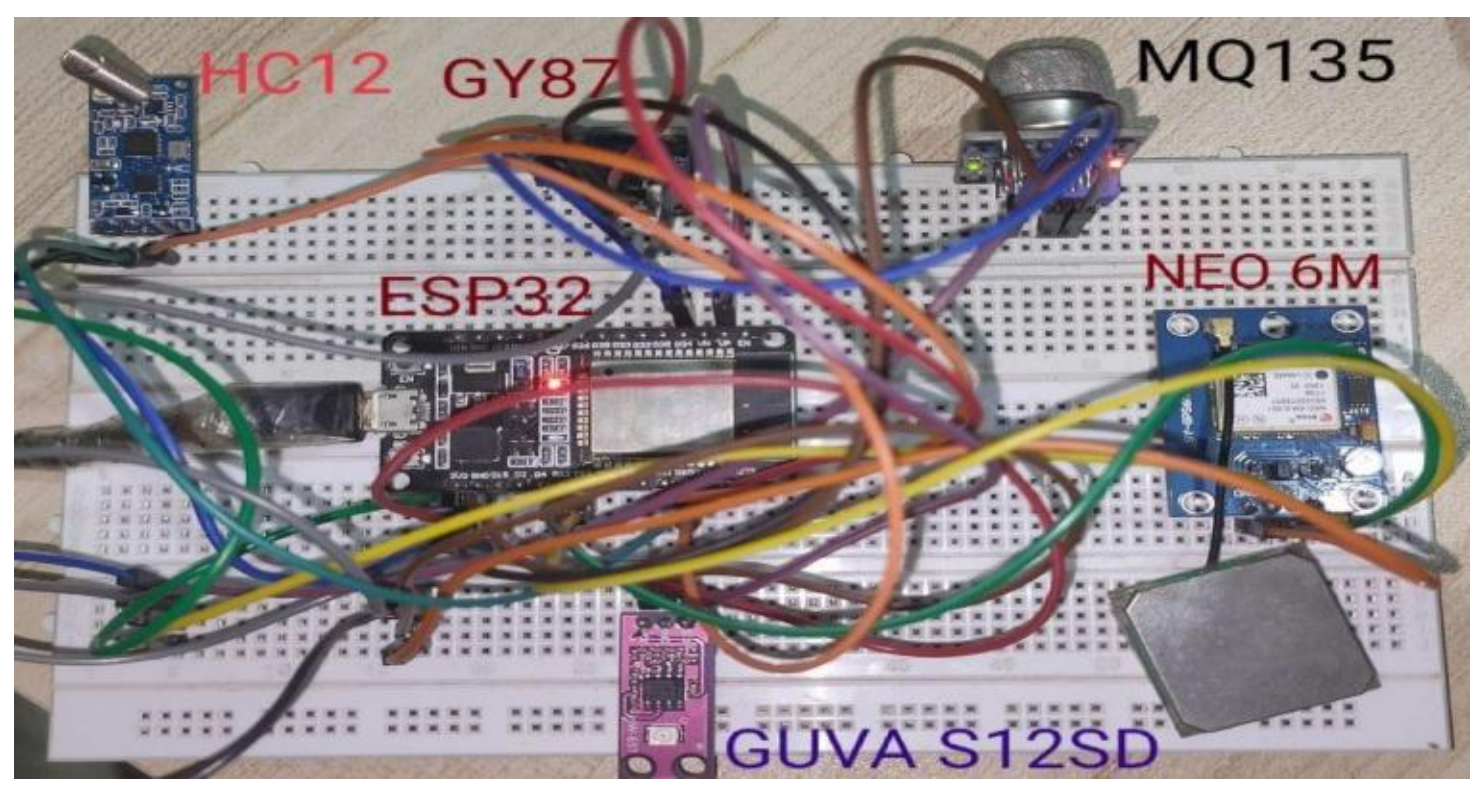

Figure 7: Testing of an OBC for a CanSat

```
COM5

WiFi connected..!
Got IP: 192.168.43.80
HTTP server started
Onboard Satellite Subystem Started
Initialize BMP180 on GY 87
BMP180 init success
Initialize MPU6050 on GY 87

 * Sleep Mode:            Disabled
 * Clock Source:          PLL with X axis gyroscope reference
 * Accelerometer:         +/- 2 g
 * Accelerometer offsets: 2736 / -1695 / 1112

¿
**************** UV SENSOR - GUVA S12SD READING ****************
sensor reading = 335.00
sensor voltage = 0.27 V
UV Index: 2
**************** AIR QUALITY SENSOR - MQ135 READING ****************
AirQua=83 PPM

**************** TEMP+HUM+ALT SENSOR - BMP180 READING ****************
Provided altitude: 50 meters, 164 feet
Temperature: 32.66 deg C, 90.79 deg F
Absolute Pressure: 999.37 mb, 0.00 inHg
Relative (sea-level) pressure: 0.00 mb, 0.00 inHg
Computed Altitude: nan meters, nan feet

**************** GYROSCOPE SENSOR - MPU6050 READING ****************
 Pitch = 0.00 Roll = -0.00 Yaw = 0.00

**************** ACCELEROMETER SENSOR - MPU6050 READING ****************
 Xraw = 17324.00 Yraw = 65500.00 Zraw = 61608.00
 Xnorm = 10.32 Ynorm = 0.01 Znorm = 36.88

**************** ESP32 INPUT READINGDS ****************
Input Voltage to ESP32: 5 V Input Current to ESP32: 128 mA
--------------------------------------------------------------------------
```

Figure 8: Sensor Data received from OBC to Ground Station

192.168.43.80

**Onboard Computer System for a CanSat**

**Real-time Data:**

Airquality: 94 PPM

UV Index: 3

UV Sensor Value: 447

Current Altitude: 49.2 m

Temperature: 31.43 °C

Absolute Pressure: 1013.59 mB

Pitch: -17.89 Roll: 0.04 Yaw: 10.25

X acc.: 8.63

Y acc.: 0.01

Z acc.: 21.06

Latitude: 20.278863

Longitude: 72.878662

Input current to ESP32: 127 mA

Input Voltage to ESP32: 4.99 V

Figure 9: OBC's Data on Web Server

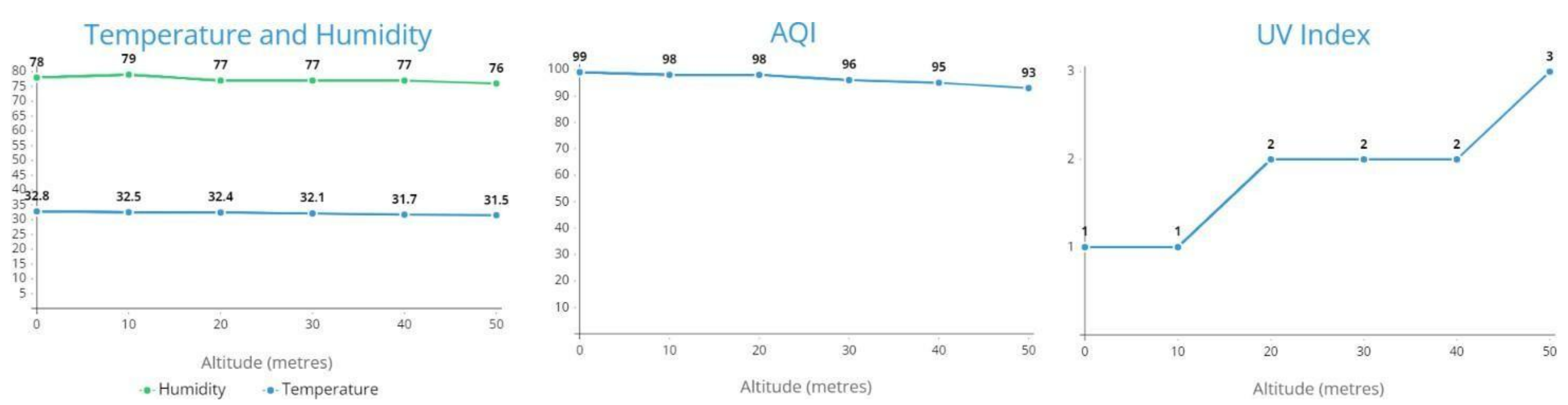

Figure 10: Changes in atmospheric parameters with respect to altitude

## 6. Conclusion

This paper presented a practical and cost-efficient solution for gathering and relaying environmental data fromhigh altitudes usingthe CanSat OBC. It comprises varioussensors to collect and transmit data, such as temperature, humidity, pressure, altitude, air quality, UV radiation, and GPS location. By integrating the OBC with a web server, real-time access to data was made possible, enabling more in-depth analysis and informed decision-making. Nevertheless, the OBC can benefit from further improvementsto enhance its accuracy, processing speed, and transmission range. Possible enhancements include incorporating machine learning algorithms to optimize data processing and analysis, as well as integrating advanced wireless communication technologies to extend the reach of data transmission. Overall, this study lays the groundwork for future research and development of environmental monitoring from high altitudes.

## 7. Recommendations

As technology continues to advance, future upgrades to the OBC can further enhance its capabilities and expand its potential uses beyond its current functionality. The low-cost OBC used in a CanSat can serve as a valuable tool for mission control.

The OBC of a CanSat can be improved in various ways to enhance its performance, including using low-power and highly efficient boards and FPGAs. In addition to these factors, other considerations, such as the selection of a better microcontroller and sensors, can also play a significant role in determining the optimal Onboard Computer System for a specific application. With the ability to adapt to new technologies, the OBC has the potential to become an even more integral component in the success of CanSat missions.